# Excitation functions of proton induced nuclear reactions on $^{nat}$W up to 40 MeV

M. U. Khandaker [a], M. S. Uddin [a,b], K. S. Kim [a], M. W. Lee [a], Y. S. Lee [a], G. N. Kim [a1]

[a] *Department of Physics, Kyungpook National University, Daegu 702-701, Korea*

[b] *Institute of Nuclear Science and Technology, Atomic Energy Research Establishment, Savar, GPO Box No.3787, Dhaka-1000, Bangladesh.*

**Abstract**

Excitation functions for the production of the $^{181,182m,182g,183,184g,186}$Re and $^{183,184}$Ta radionuclides from proton bombardment on natural tungsten were measured using the stacked-foil activation technique for the proton energies up to 40 MeV. A new data set has been given for the formation of the investigated radionuclides. Results are in good agreement with the earlier reported experimental data and theoretical calculations based on the ALICE-IPPE code. The thick target integral yields were also deduced from the measured excitation functions. The deduced yield values were compared with the directly measured thick target yield (TTY), and found acceptable agreement. The investigated radionuclide $^{186}$Re has remarkable applications in the field of nuclear medicine, whereas the data of $^{183,184g}$Re and $^{183}$Ta have potential applications in thin layer activation analysis and biomedical tracer studies, respectively.

**1. Introduction**

Cross-sections data of particle induced activation on tungsten (W) are of interest in various fields like model calculations, accelerator technology, charged particle activation analysis, medical radioisotope production, thin layer activation analysis to wear, corrosion in machine components etc.

---

[1] Corresponding author. Tel.: +82-53-950-5320, Fax :+82-53-955-5356
E-mail address: gnkim@knu.ac.kr

Reactor- and accelerator-produced radionuclides are widely used in nuclear medicine and represent powerful tools in diagnostic and therapeutic procedures [1]. The information on the excitation functions of residual nuclei are also important for verification of different models used to explain the reaction mechanism, optimization of the production yield, and to estimate the radionuclide impurities of simultaneously produced radionuclides.

Tungsten is an ideal target material for the production of medically important radioisotopes ($^{186}$Re, $^{188}$Re, $^{186}$W/$^{186}$Re etc.). Rhenium radionuclides have varieties of application in nuclear medicine including radioimmunotherapy, radionuclide synovectomy, and bone pain palliation [2]. The most widely used rhenium radioisotopes $^{186}$Re and $^{188}$Re have suitable properties for radiotheraphy. Neutron rich radionuclide $^{186}$Re can be produced through two routes; by reactor through (n,γ) process, and by cyclotron through $^{186}$W(p,n)$^{186}$Re process. The later is better due to its carrier-free nature and also high specific activity. High specific activity is generally required for radiolabeling of tumor-specific antibodies [3]. Recent survey has shown that $^{186}$Re is an ideal candidate for radioimmunotherapy [4-6] of its moderate $\beta^-$ particle energies at 1.07 and 0.933 MeV, low- abundance (9%) γ emission at 137 keV, which allows for in vivo tracking of the radiolabeled biomolecules and estimation of dosimetry calculation. The suitable 3.7-day half-life allows sufficient time for the synthesis and shipment of potential radiopharmaceuticals.

Several investigations [7-9] were carried out for the production of high specific activity $^{186}$Re radionuclide in no carrier added (NCA) form by proton bombardment on enriched tungsten targets using cyclotron, but large discrepancies are found among them. On a practical level, it is very difficult to estimate the causes of discrepancies among the data sets. These inconsistencies severely limit the reliability of data evaluations. Until now, no recommended data are available for the mentioned ($^{186}$Re) radionuclide. However, nowadays only a few groups are engaged with the systematic investigations of (p,xn) processes leading to the production of medically important radionuclide by using hospital based small sized cyclotrons. The aim of the present work was to report reliable excitation functions for the $^{nat}$W(p, xn) nuclear reactions in the energy region up to 40 MeV, and hence to increase the reliability of the literature database.

## 2. Experimental

### 2.1. Targets and Irradiations

Excitation functions for the proton-induced reactions on natural tungsten were measured by using the well established stacked foil activation technique. High purity (>99.99%) metallic form of tungsten ($^{180}$W-0.13%, $^{182}$W-26.3%, $^{183}$W-14.3%, $^{184}$W-30.67%, $^{186}$W-28.6%), copper and aluminum foils with natural isotopic compositions were assembled in the stacks. Two stacks were arranged to perform the individual experiments. Al and Cu foils were used as monitor as well as energy degraders. An Al foil (509 μm thickness) was placed at the front of each stack, where the beam energy is well defined, to accurately measure the beam flux. Special care was taken in preparation of uniform targets with known thickness, determination of proton energy and intensity along the stacks. The precisely measured thickness of the W, Cu, and Al foils were 203 μm, 50 μm and 100 μm, respectively. The thickness of the foils was calculated on the basis of weight measurement of the whole sheet of foil. Two stacks were separately irradiated for 0.5 hour by the proton energy of 42 MeV with a beam current of about 100 nA from the external beam line of the MC50 cyclotron at the Korea Institute of Radiological and Medical Sciences (KIRAMS). The beam intensity was kept constant during the irradiation. It was necessary to ensure that equal areas of monitor and target foils intercepted the beam. The irradiation geometry was kept in a position so that the foils get the maximum beam line. To avoid errors in the determination of the beam intensity and energy, excitation functions of the monitor reactions were remeasured simultaneously with the reactions induced on natural tungsten.

*2.2. Measurement of Radioactivity*

The activities of the produced residual radionuclides were measured non-destructively by using high purity germanium (HPGe) γ-ray spectrometry coupled to ORTEC (PopTop, Gmx20) NIM-based high bias voltage, linear amplifier, analogue to digital converter (ADC) and PC-based multi-channel analyzer (MCA). The activity measurements of the irradiated samples were started about 15 h after the end of the bombardment (EOB). This cooling time was enough to the complete decay of most of the undesired short-lived nuclides so that we could easily identify and separate the complex gamma lines. Conversely, due to this cooling time the peak in spectrum for short-lived isotopes was lost. The counting of each sample was done 3-4 times by giving long time interval to avoid disturbance by overlapping gamma-lines from undesired sources, and also more precisely assay the longer-lived radionuclides. Interactive peak analysis was done using the program Gamma Vision (*EG&G ORTEC*). All samples were counted at distances of 15 cm and 25 cm from the end-cap of the detector to avoid coincidence losses, to assure low dead time (<10%) and point like geometry.

The efficiency versus energy curves of the HPGe-detector for the counting distances were determined using standard gamma-ray point sources, $^{133}$Ba, $^{109}$Cd, $^{22}$Na, $^{60}$Co, $^{57}$Co, $^{54}$Mn and $^{137}$Cs. The proton beam intensity was determined from the measured activities induced in aluminum and copper monitor foils at the front position of each stack using the reactions $^{27}$Al(p,x)$^{22,24}$Na and $^{nat}$Cu(p,x)$^{62}$Zn [10], respectively. The use of several monitor foils decreases the probability of introducing unknown systematic errors in activity determination. The beam intensity was considered constant to deduce cross sections for each foil in the stack. The loss of proton energy for each foil in the stack was calculated by using the computer program SRIM-2003 [11].

The activation cross-sections for the reactions $^{nat}$W(p, xn) were determined in the proton energy range 6-40 MeV by using the well-known activation formula [12]. The decay data of the radioactive products were taken from the NUDAT database (*National Nuclear Data Center, information extracted from the NuDat database, http://www.nndc.bnl.gov/nudat2*), and are collected in Table 1. The threshold energies given in Table 1 were taken from the Los Alamos National Laboratory, T-2 Nuclear Information Service on the internet [13].

In the present experiment, all the errors were considered as independent. Consequently, they were quadratically added according to the laws of error propagation to obtain total errors. Moreover, some of the sources of errors are common to all data, while others individually affect each reaction. However, the combined uncertainty in each cross section was estimated by considering the following uncertainties; statistical uncertainty of gamma-ray counting (0.5-10 %), uncertainty in the monitor flux (~7 %) and uncertainty in the efficiency calibration of the detector (~4 %). The overall uncertainties of the cross section measurements were in the range 8-13 %.

## 3. Results and discussion

In natural tungsten, there are five stable isotopes which can transmute to corresponding the Re and Ta radioisotopes by the W(p,xn)Re and W(p,pxn)Ta reactions, respectively. In some cases, two or more γ-rays were used for the measurement of each reaction cross-section, and the average value is presented. The excitation functions of the investigated radioactive products $^{181,182g,182m,183,184g,186}$Re and $^{183,184}$Ta are shown in Figs. 1–8. The numerical data with errors are collected in Table 2-3. The integral yields were deduced using the measured cross-sections taking into account that the total energy is absorbed in the targets and are shown in Figs. 9–14.

*3.1. Cross-sections of residual radioisotopes of Rhenium*

### 3.1.1. Production of $^{181}Re$

The radioisotope $^{181}$Re ($T_{1/2}$= 19.9 h) has no isomeric state. The open reaction channels for the direct production of $^{181}$Re radionuclide have shown in Table 1. Excitation function for the formation of the $^{181}$Re radionuclide is shown in Fig. 1 together with the literature values. Only a few experimental data are available in the literature for the investigated energy region. Recently, Lapi et al. [14] and Zhang et al. [15] reported activation cross-sections data for rhenium radioisotopes up to 17.6 MeV and 25 MeV, respectively. Where as, Tarkanyi et al. [16] reported production cross section data of Re and Ta radionuclides up to 34 MeV. An excellent agreement is found with Lapi et al. [14] and Zhang et al. [15] data, while Tarkanyi et al. [16] and the ALICE-IPPE model [17] calculation appear systematically higher values than this work.

### 3.1.2. Production of $^{182g}Re$

The radioisotope $^{182}$Re has two longer-lived energy states. The metastable state radioisotope $^{182m}$Re ($T_{1/2}$=12.7 h) has no internal transition to the ground state ($T_{1/2}$=64.0 h) and the decay of both states is accompanied by different number of γ-lines. Hence, we could measure the production cross-sections for both states independently. The produced activity from $^{182g}$Re was measured using the independent gamma lines (229.32, 1076.2 and 1231.0 keV) and found consistent result among them. Excitation function for the $^{182g}$Re radionuclide production is presented in Fig. 2 together with the available literatures, as well as ALICE-IPPE code [17]. We have found an excellent agreement with the data reported by Lapi et al. [14] and Zhang et al. [15]. Good agreement was also found with Tarkanyi et al. [16] data in both values and trend of the peak formation. Szelecsenyi et al. [18] also reported the production cross section data of $^{182g}$Re radionuclide up to 18 MeV proton energy. The agreement with Szelecsenyi et al. [18] data is reasonably good up to 11 MeV, but after then their [18] data are systematically lower than our new values, as well as other literature values. The trend of peak formation of our data agrees with the theoretical data of ACILE-IPPE code [17]. Actually, this code gives the sum of cross sections for the formation of $^{182g}$Re and $^{182m}$Re.

### 3.1.3. Production of $^{182m}Re$

The isomeric state $^{182m}$Re has only one independent gamma line 470.32 keV ($I_\gamma$=2 %) with low intensity. The other high intense gamma-rays 1121.4, 1189.2 and 1221.5 keV have interfering contribution from $^{182}$Ta radioisotope. In principle, the activity of this isomeric state can be determined by decomposition of strong common gamma-lines by measuring the decay curve. Comparing the present experimental conditions (i.e., irradiation time, decay time, measuring time) with the half life of $^{182}$Ta (T$_{1/2}$= 114.43 d), we have considered that the contribution from $^{182}$Ta in the common gamma lines is very negligible amount. However, we have confirmed the consistency of measured values by using less intense independent gamma line 470.32 keV. The present experimental cross section data of $^{182m}$Re is shown in Fig. 3. The present results were compared with the available literature values, as well as theoretical values of ALICE-IPPE code [17]. A good agreement was found with the literature data reported by Lapi et al. [14] and Tarkanyi et al. [16] in both shape and size. However, the data reported by Zhang et al. [15] showed systematically lower values, but the trend of peak formation agrees with our new results.

### 3.1.4. Production of $^{183}$Re

The $^{183}$Re (T$_{1/2}$=70 d) radionuclide has no longer-lived isomeric state. The produced activity of $^{183}$Re was determined using the independent gamma line 162.32 keV. The measured excitation function for the production of $^{183}$Re radionuclide and the literature data are shown in Fig. 4. The present data are in good agreement with Tarkanyi et al. [16], except only one point at proton energy 15.9 MeV. The data reported by Zhang et al. [15] showed little lower values. Data of Schoen et al. [7] and Lapi et al. [14] showed significantly lower and higher values, respectively, than our new experimental data. Since the experimental technique and data evaluation method are more or less same to all groups, the use of monitor cross sections data may be the key point to get scattered data. Since the trend of peak formation of Schoen et al. [7] data agrees with the latest literature values, it could be corrected by using accurate monitor cross sections data.

### 3.1.5. Production of $^{184}$Re

The radioisotope $^{184}$Re has two longer-lived isomeric states. The metastable state $^{184m}$Re (T$_{1/2}$=169 d) mainly decays to the ground state (T$_{1/2}$=38 d) by internal transition (75.4%). As the half-life of the metastable state is four times longer than that of the ground state, contributions from the decay of $^{184m}$Re to $^{184g}$Re strongly depend on the period of the activity measurement of $^{184g}$Re. In principle, the production of the two states can be measured by assessment of different independent gamma-lines following the decays of the two isomers. However, we measured the production cross section of $^{184g}$Re by using the independent gamma line 894.76 keV (I$\gamma$ = 15.6 %). Since in the present experiment, the irradiation time (t$_{irr}$= 0.5 h), decay time (t$_d$= few hours) and the measuring time (t$_m$= 200 ~1000 sec) are negligible compare to the half life of metastable state (T$_{1/2}$=169 d), hence we could considered that the gamma lines originating from the decay of the ground state contain only an insignificant contribution from the metastable state. The cross-section for the direct production of the ground state could hence be determined with very small additional uncertainty. The measured production cross sections of $^{184}$Re radionuclide and the available experimental data are shown in Fig. 5. An excellent agreement has found with all the experimental data reported by Lapi et al. [14], Zhang et al. [15], Tarkanyi et al. [16], Schoen et al. [7], and also with the theoretical data from ALICE-IPPE code [17].

### 3.1.6. Production of $^{186}$Re

The most suitable reaction path for the production of $^{186}$Re from the proton bombardment on natural tungsten is $^{186}$W(p, n)$^{186}$Re (Q= -1.069 MeV). The measured excitation function for the production of neutron rich therapeutic radionuclide $^{186}$Re has shown in Fig. 6 together with the available experimental values and theoretical data taken from ALICE-IPPE code [17]. We found a very good agreement with the data reported by Lapi et al. [14], Tarkanyi et al. [16], Szelecsenyi et al. [18], and Shigeta et al. [8] in the investigated energy region. The data reported by Zhang et al. [15] and ALICE-IPPE prediction [17] showed lower values in the peak formation region at 6-13 MeV.

### 3.2. Cross-sections of residual radioisotopes of Tantalum
### 3.2.1. Production of $^{183}$Ta

The longer-lived $^{183}$Ta radionuclide offers considerable advantages as a biomedical tracer. $^{183}$Ta has suitable nuclear decay properties with a half-life of 5.1 days; and decay by β⁻ radiation with accompanying γ-radiation at 246keV (27%), 353 keV (11.2%), and 108 keV (11.9%). This $^{183}$Ta (7/2$^+$) radioisotope could be produced directly via (p,px) reactions on W or through the decay of simultaneously formed Re radionuclide. At low energies the α-particle emission dominates while at higher energies the individual particle emission is more significant. Only one earlier data reported by Tarkanyi et al. [16] was found in the literature. A considerable disagreement was found (Fig. 7) in the energy range 21-31 MeV between the present work and Tarkanyi et al. [16] data. We couldn't determine any potential reason for this disagreement. In the ALICE-IPPE calculation [17], probably the α particle emission reaction was not considered, hence predicting cross-sections are at very lower values.

### 3.2.2. Production of $^{184}$Ta

The reaction channel for the production of $^{184}$Ta (8.70 h) radioisotope from the proton bombardment on natural tungsten is $^{186}$W(p,n2p). The produced activity was determined using the intense independent gamma line 414.01 keV. For this process no earlier experimental results were found in the literature. The excitation function for this radioisotope production is shown in Fig. 8.

3.3. Integral yields

The integral yields for the production of medically and technologically relevant activation products (Re, Ta) were deduced using the experimental cross-sections and stopping power of $^{nat}$W over the energy range from threshold up to 40 MeV. It is expressed as MBq µA$^{-1}$h$^{-1}$, i.e. for an irradiation at beam current of 1 µA for 1 hour. The obtained results are shown in Figs. 9-14 as a function of proton energy in comparison with the few directly measured thick target yields (TTY) found in the literature. Dmitriev and Molin [19,20] measured integral yield as a function of the energy up to 22 MeV for $^{181,182m,182g,183,184m,184g}$Re. The comparison of our calculated data and the experimental yield data showed a good agreement for $^{181}$Re, $^{183}$Re, and $^{184g}$Re, but the yields differ significantly for production of $^{182m}$Re, and $^{182g}$Re; however the area of their ($^{182m}$Re, $^{182g}$Re) application is limited.

**5. Conclusions**

Owing to the original goal, a new data set for the production cross sections of the $^{181,182m,182g,183,184m,184g,\ 186}$Re and $^{183,184}$Ta radionuclides through the proton bombardment on natural tungsten have been reported in the energy range of 6.61-40 MeV using the stacked-foil activation technique with an overall uncertainty of about 13%. The measured cross section data for most of the radionuclides have much importance in several fields; e.g., nuclear medicine, thin layer activation process, trace element analysis, and for the improvement of model calculations.

Rhenium-186 ($^{186}$Re) is one of the most useful radionuclide for internal radiotherapy in nuclear medicine. The measured production cross sections data of this radionuclide will help effectively to optimize the production conditions. The production cross sections data of $^{184g}$Re and $^{183}$Re radioisotopes have much importance in thin layer activation technique due to their convenient half-lives, gamma-lines and high yields. The longer-lived $^{183}$Ta radionuclide could be used as a biomedical tracer due to its suitable nuclear decay properties; hence the production cross section of this radionuclide has much significance. The measured production cross sections data of the other $^{181}$Re, $^{182m}$Re, $^{182g}$Re, and $^{184}$Ta radionuclides could help in the prediction, optimization and evaluation of the radiochemical purity. Furthermore, it should be mentioned that for precise estimation of the impurity levels in fully and/or partly enriched targets, isotopic cross-sections data are required for all stable isotopes of that target material. At present the isotopic cross-sections data of tungsten are not sufficient in our investigated energy range, hence these new data have importance in this regard also.

Although, most of the medically important radionuclides are produced commercially by using nuclear reactors through (n,γ) process, the cyclotron offers the alternative production route of these isotopes in no carrier added form (NCA) with high specific activity. In the last two decades, the rapid installations of hospital based cyclotrons all over the world were driven by the advent of advances in PET imaging technique. However, the measured production cross sections of $^{nat}$W(p,xn) process would be very helpful to prepare a recommended data base leading to various mentioned applications.

**Acknowledgements**

The author would like to express their sincere thanks to the staffs of the MC50 Cyclotron Laboratories for their cordial help in performing the experiment. This work is partly supported through Project number M20602000001-06B0200-00110 of the Ministry of Science and Technology (MOST) and through the Science Research Center (SRC) program of the Institute of High Energy Physics, Kyungpook National University.

**References**

[1]   G.B. Saha, Fundamentals of nuclear pharmacy. 3rd Ed. New York: Springer-Verlag; 1992, p. 3145.
[2]   L.F. Mausner, S.C. Srivastava, Med. Phys. 20 (1993) 503.
[3]   W.A. Volkert, W.F. Goeckeler, G.J. Ehrhardt, A.R. Ketring, J. Nucl. Med. 32 (1991) 174.
[4]   S. Kinuya, K. Yokoyama, M. Izumo, T. Sorita, T. Obata, H. Mori, K. Shiba, N. Watanabe, N. Shuke, T. Michigishi, N. Tonami, Cancer Lett. 219 (2005) 41.
[5]   S. Kinuya, K. Yokoyama, M. Izumo, T. Sorita, T. Obata, H. Mori, K. Shiba, N. Watanabe, N. Shuke, T. Michigishi, N. Tonami, J. Cancer Res. Clin. Oncol. 129 (2003) 392.
[6]   E.J. Postema, P.K. Borjesson, W.C. Buijs, J.C. Roos, H.A. Marres, O.C. Boerman, R. de Bree, M. Lang, G. Munzert, G.A.V. Dongen, W.J. Oyen, J. Nucl. Med. 44 (2003)1690.
[7]   N.C. Schoen, G. Orlov, R.J. McDonald, Phys. Rev., Part C 20 (1979) 88.

[8] N. Shigeta, H. Matsuoka, A. Osa, M. Koizumi, M. Izumo, K. Kobayashi, K. Hashimoto, T. Sekine, R.M. Lambrecht, J. Radioanal. Nucl. Chem. 85 (1996) 205.

[9] E.M. Moustapha, G.J. Ehrhardt, C.J. Smith, L.P. Szajek, W.C. Eckelman, S.S. Jurisson, J. Nucl. Med. Biol. 33 (2006) 81.

[10] F. Tarkanyi, S. Takacs, K. Gul, A. Hermanne, M.G. Mustafa, M. Nortier, P. Oblozinsky, S. M. Qaim, B. Scholten, Yu. N. Shubin, Z. Youxiang, IAEA-TECDOC-1211, Beam monitor reactions, Charged Particle Cross-Section Database for Medical Radioisotope Production: Diagnostic Radioisotopes and Monitor Reactions, Co-ordinated Research Project (1995-1999). IAEA, Vienna. Austria, May 2001. Available from <http://www-nds.iaea.org/medical/>.

[11] J.F. Ziegler, J.P. Biersack, U. Littmark, SRIM 2003 code, Version 96.xx. The stopping and range of ions in solids. Pergamon, New York. <http://www.srim.org/>

[12] G. Gilmore, J.D. Hemingway, Practical Gamma-Ray Spectrometry, John Wiley & Sons, England, Chapter 1, 1995, p. 17.

[13] Reaction Q-values and thresholds, Los Alamos National Laboratory, T-2 Nuclear Information Service. Available from< http://t2.lanl.gov/data/qtool.html>

[14] S. Lapi, W.J. Mills, J. Wilson, S. McQuarrie, J. Publicover, M. Schueller, D. Schyler, J.J. Ressler, T.J. Ruth, App. Radiat. Isot. 65 (2007) 345.

[15] X. Zhang, W. Li, K. Fang, W. He, R. Sheng, D. Ying, W. Hu, Radiochim. Acta 86 (1999) 11.

[16] F. Tarkanyi, S. Takacs, F. Szelecsenyi, F. Ditroi, A. Hermanne, M. Sonck, Nucl. Instr. and Meth. B 252 (2006) 160.

[17] Yu.N. Shubin, V.P. Lunev, A.Yu. Konobeyev, A.I. Dityuk, MENDL-2P, Proton reaction data library for nuclear activation (Medium Energy Nuclear Data Library), IAEA-NDS-204, 1998.

[18] F. Szelecsenyi, S. Takacs, F. Tarkanyi, M. Sonck, A. Hermanne, in: J.R. Heys, D.G. Mellilo, Chichester, et al. (Eds.), Proceedings of the Sixth International Symposium on Synthesis and Applications of Isotopically Labelled Compounds, Philadelphia, USA, 14–18 September, 1997, John Wiley and Sons, 1998, p. 701

[19] P.P. Dmitriev, G.A. Molin, Atom. Energy. 48 (1980) 122.

[20] P.P. Dmitriev, G.A. Molin, Yadernie Konstanti 44 (5) (1981) 43, Report INDC(CCP)-188/L 1983.

**Figure captions:**

Fig. 1 Excitation function of the $^{nat}W(p, x)^{181}Re$ reaction
Fig. 2 Excitation function of the $^{nat}W(p, x)^{182g}Re$ reaction
Fig. 3 Excitation function of the $^{nat}W(p, x)^{182m}Re$ reaction
Fig. 4 Excitation function of the $^{nat}W(p, x)^{183}Re$ reaction
Fig. 5 Excitation function of the $^{nat}W(p, x)^{184g}Re$ reaction
Fig. 6 Excitation function of the $^{nat}W(p, x)^{186}Re$ reaction
Fig. 7 Excitation function of the $^{nat}W(p, x)^{183}Ta$ reaction
Fig. 8 Excitation function of the $^{nat}W(p, x)^{184}Ta$ reaction
Fig. 9 Integral yields for the production of the $^{181}Re$ radionuclide
Fig. 10 Integral yields for the production of the $^{182m, 182g}Re$ radionuclides
Fig. 11 Integral yields for the production of the $^{183}Re$ radionuclide
Fig. 12 Integral yields for the production of the $^{184g}Re$ radionuclide
Fig. 13 Integral yields for the production of the $^{186}Re$ radionuclide
Fig. 14 Integral yields for the production of the $^{183, 184}Ta$ radionuclides

Table 1
Decay data of the produced radionuclide from natural tungsten

| Nuclide | Half-life | $E_\gamma$ (keV) | $I_\gamma$ (%) | Contributing reactions | Q-value (MeV) | Threshold (MeV) |
|---|---|---|---|---|---|---|
| $^{181}$Re | 19.9 h | 360.70 | 20.0 | $^{182}$W(p,2n) | -10.58 | 10.64 |
| | | **365.57** | **57.0** | $^{183}$W(p, 3n) | -16.78 | 16.87 |
| | | 639.30 | 6.4 | $^{184}$W(p,4n) | -24.19 | 24.32 |
| | | 805.2 | 3.1 | $^{186}$W(p,6n) | -37.14 | 37.34 |
| | | 953.42 | 3.6 | | | |
| | | 1000.2 | 3.3 | | | |
| $^{182m}$Re | 12.7 h | 470.32 | 2.0 | $^{182}$W(p,n) | -3.58 | 3.60 |
| | | 1121.4 | 31.8 | $^{183}$W(p,2n) | -9.77 | 9.83 |
| | | **1189.2** | **15** | $^{184}$W(p,3n) | -17.18 | 17.28 |
| | | 1221.5 | 24.8 | $^{186}$W(p,5n) | -30.13 | 30.30 |
| $^{182g}$Re | 64.0 h | 169.15 | 11.3 | $^{182}$W(p,n) | -3.58 | 3.60 |
| | | **229.32** | **25.7** | $^{183}$W(p,2n) | -9.77 | 9.83 |
| | | 256.45 | 9.5 | $^{184}$W(p,3n) | -17.18 | 17.28 |
| | | 276.31 | 8.7 | $^{186}$W(p,5n) | -30.13 | 30.30 |
| | | 286.56 | 7.0 | | | |
| | | 357.07 | 10.3 | | | |
| | | **1076.2** | **10.3** | | | |
| | | 1121.3 | 22 | | | |
| | | 1221.4 | 17.4 | | | |
| | | **1231.0** | **14.9** | | | |
| | | 1427.3 | 9.8 | | | |
| $^{183}$Re | 70.0 d | **162.32** | **23.3** | $^{183}$W(p,n) | -1.34 | 1.35 |
| | | 291.72 | 3.05 | $^{184}$W(p,2n) | -8.75 | 8.80 |
| | | 354.0 | 0.54 | $^{186}$W(p,4n) | -21.70 | 21.82 |
| $^{184m}$Re | 169 d | 104.73 | 13.4 | $^{184}$W(p,n) | - 2.27 | 2.28 |
| | | 161.27 | 6.5 | $^{186}$W(p,3n) | -15.21 | 15.29 |
| | | **216.55** | **9.4** | | | |
| | | 792.06 | 3.69 | | | |
| $^{184g}$Re | 38.0 d | 792.07 | 37.5 | $^{184}$W(p,n) | - 2.27 | 2.28 |
| | | **894.76** | **15.6** | $^{186}$W(p,3n) | -15.21 | 15.29 |
| | | 903.28 | 37.9 | | | |
| $^{186}$Re | 3.72 d | 137.16 | 9.42 | $^{186}$W(p,n) | -1.36 | 1.37 |
| $^{183}$Ta | 5.1 d | **246.06** | **27.0** | $^{186}$W(p,α) | 7.65 | 0.0 |
| | | 353.99 | 11.2 | $^{184}$W(p,2p) | - 7.70 | 7.74 |
| $^{184}$Ta | 8.70 h | 215.34 | 11.4 | $^{186}$W(p,n2p) | -15.03 | 15.11 |
| | | **318.04** | **22.8** | | | |
| | | 384.28 | 12.5 | | | |
| | | **414.01** | **72** | | | |
| | | 792.07 | 14.5 | | | |
| | | 903.29 | 15 | | | |

Table 2

Measured cross-sections of the reactions $^{nat}W(p,xn)^{181, 182g, 182m, 183}Re$

| Proton Energy | Cross sections with uncertainties (mb) | | | |
|---|---|---|---|---|
| MeV | $^{181}$Re | $^{182g}$Re | $^{182m}$Re | $^{183}$Re |
| 6.61 | -- | 7.64±0.97 | 27.52±2.17 | 13.36±2.11 |
| 10.80 | 18.98±3.39 | 8.56±1.14 | 102.22± 9.76 | 46.65±5.23 |
| 13.00 | 157.12±12.98 | 37.40±3.07 | 170.99±12.61 | 146.45±11.38 |
| 15.90 | 253.78±24.40 | 46.89±4.63 | 167.22± 15.59 | 202.63±16.92 |
| 18.95 | 333.09±26.04 | 72.98±5.72 | 213.31± 15.76 | 259.42±20.35 |
| 21.59 | 316.68±30.81 | 81.15±7.10 | 271.18± 25.67 | 152.03±15.15 |
| 23.44 | 284.32±23.15 | 108.47±8.44 | 343.65± 25.74 | 111.73±9.69 |
| 25.75 | 225.65±23.35 | 120.77±11.60 | 410.04±38.79 | 136.02±13.37 |
| 28.00 | 236.44±20.84 | 146.21±11.24 | 374.98±28.04 | 182.79±14.33 |
| 30.02 | 266.46±26.30 | 121.33±11.70 | 344.45± 32.71 | 254.91±24.40 |
| 31.54 | 301.81±29.25 | 103.33±8.04 | 261.26± 19.75 | 289.15±22.38 |
| 33.39 | 373.87±35.62 | 94.31±9.12 | 178.96±17.11 | 281.19±27.92 |
| 35.23 | 441.70±33.78 | 86.57±6.81 | 153.86±11.42 | 272.19±22.83 |
| 36.95 | 379.44±36.15 | 73.92±7.17 | 166.17±16.12 | 245.15±24.89 |
| 38.25 | 353.02±27.61 | 80.64±6.26 | 171.42± 12.67 | 238.61±19.37 |
| 39.86 | 329.82±31.77 | 91.27±8.82 | 189.24±17.67 | 225.75±22.00 |

Table 3

Measured cross-sections of the reactions $^{nat}W(p,xn)^{184g, 186}Re$ and $^{183, 184}Ta$

| Proton Energy | Cross sections with uncertainties (mb) | | | |
|---|---|---|---|---|
| MeV | $^{184g}$Re | $^{186}$Re | $^{183}$Ta | $^{184}$Ta |
| 6.61 | 29.66±5.43 | 2.39±0.32 | 4.46±3.50 | 0.13±0.06 |
| 10.80 | 61.57±7.97 | 16.87±1.66 | 13.86±1.84 | 0.22±0.06 |
| 13.00 | 69.70±6.58 | 11.31±0.97 | 18.56±2.78 | 1.30±0.22 |
| 15.90 | 50.01±6.00 | 10.31±1.07 | 23.25±3.84 | 1.92±0.30 |
| 18.95 | 153.19±13.61 | 9.16±0.86 | 26.02±5.07 | 2.21±0.38 |
| 21.59 | 272.40±29.27 | 9.13±0.98 | 34.36±5.59 | 2.51±0.43 |
| 23.44 | 338.94±29.57 | 7.99±0.72 | 51.65±6.54 | 1.87±0.39 |
| 25.75 | 323.30±34.78 | 7.81±0.82 | 59.41±7.33 | 1.97±0.38 |
| 28.00 | 248.55±24.62 | 9.27±0.85 | 65.48±6.62 | 1.99±0.33 |
| 30.02 | 158.85±19.81 | 7.79±0.82 | 55.61±7.92 | 2.22±0.39 |
| 31.54 | 141.50±16.29 | 7.38±0.68 | 44.25±6.69 | 2.22±0.35 |
| 33.39 | 114.56±15.71 | 7.79±0.86 | 31.72±5.23 | 2.33±0.39 |
| 35.23 | 100.31±13.82 | 7.32±0.93 | 30.60±4.73 | 2.75±0.38 |
| 36.95 | 84.14±13.84 | 7.83±0.87 | 32.35±5.22 | 2.75±0.42 |
| 38.25 | 79.47±12.06 | 7.26±0.65 | 38.67±5.67 | 2.23±0.33 |
| 39.86 | 66.43±11.63 | 6.59±0.71 | 51.12±6.61 | 1.58±0.36 |

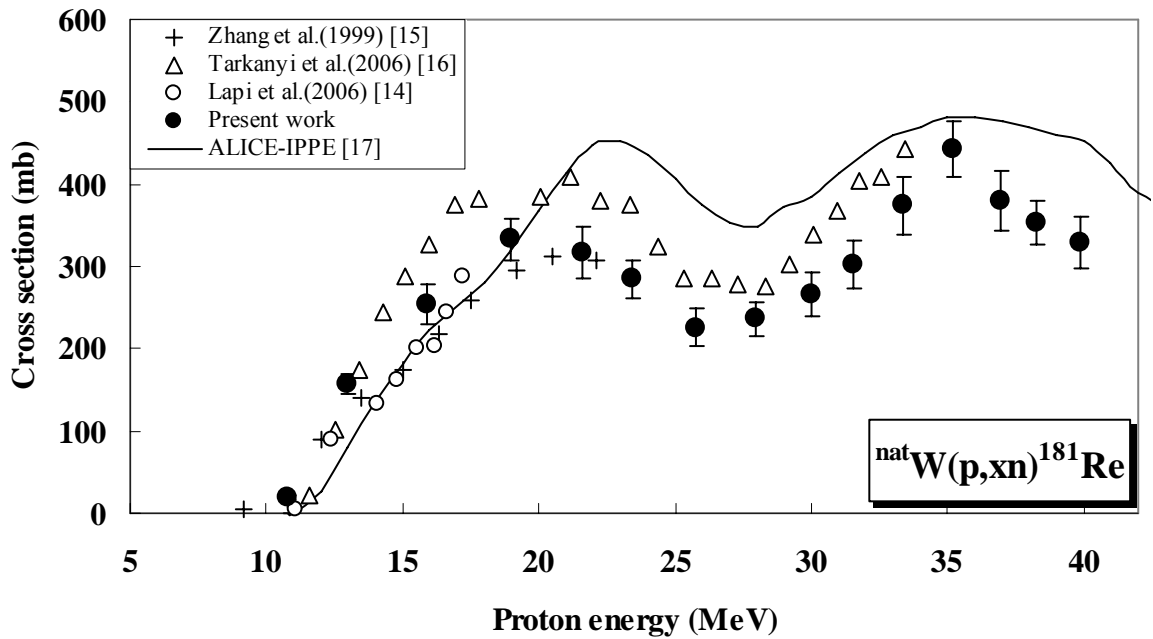

Fig. 1

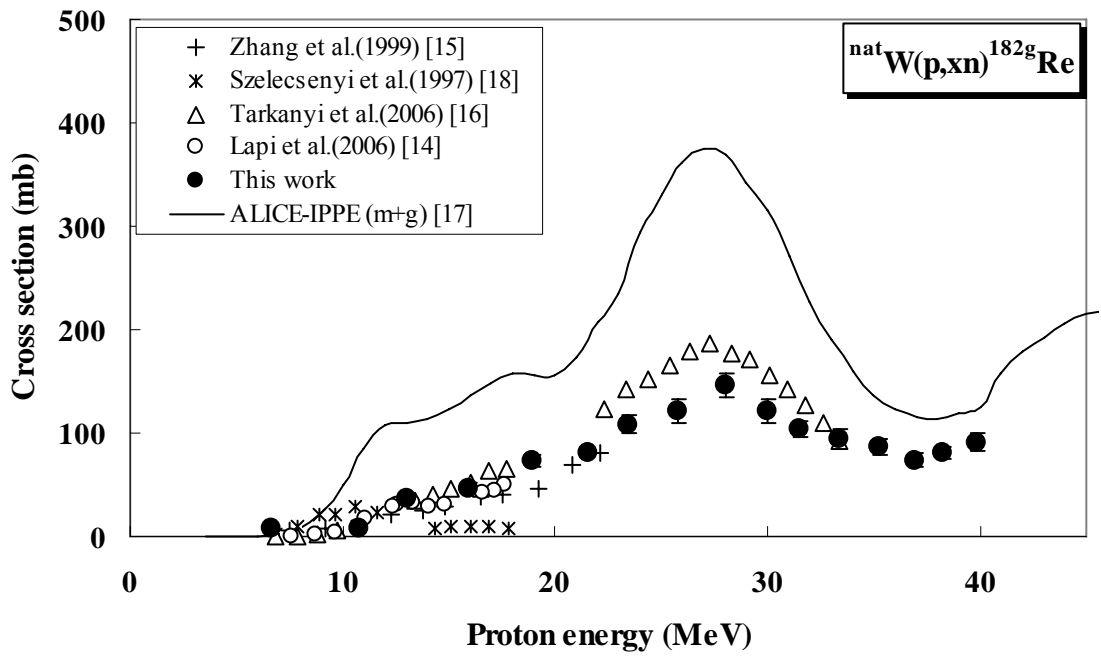

Fig. 2

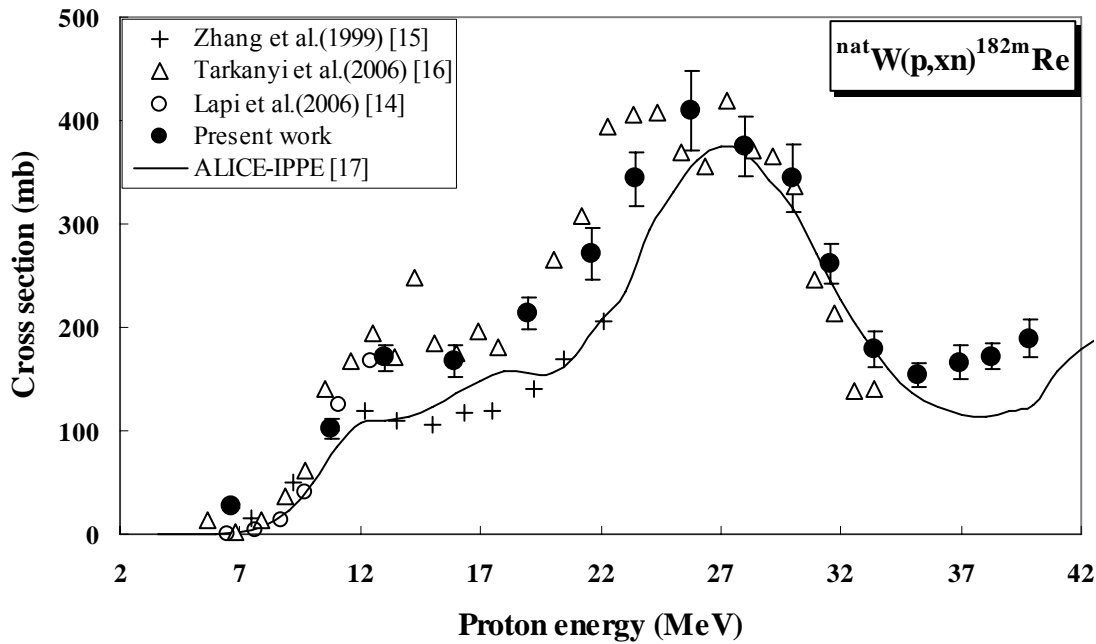

Fig. 3

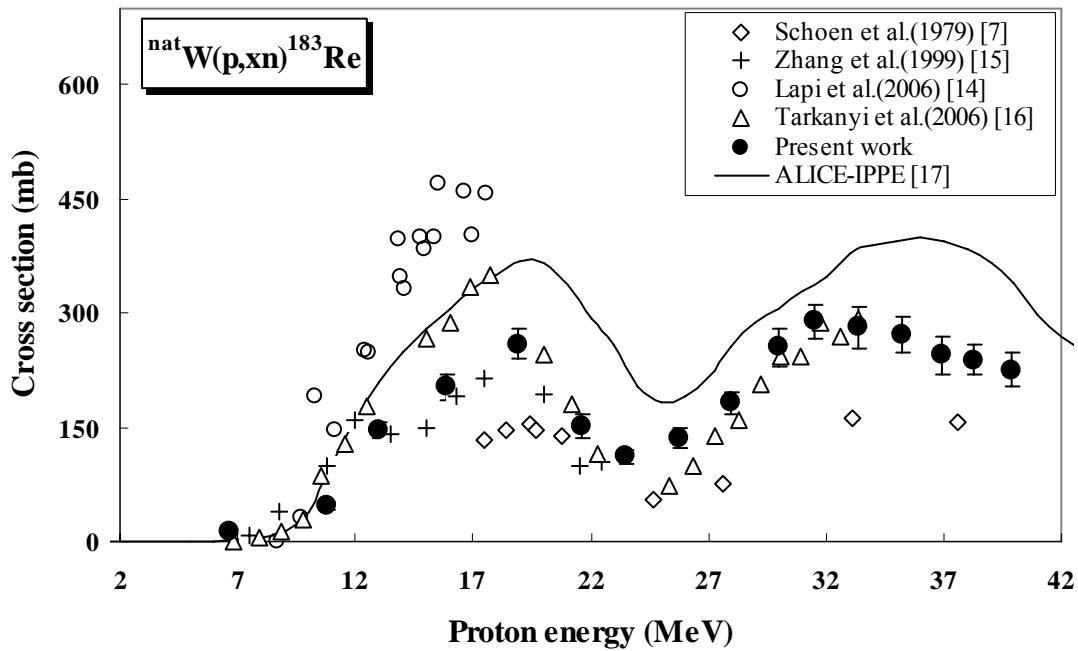

Fig. 4

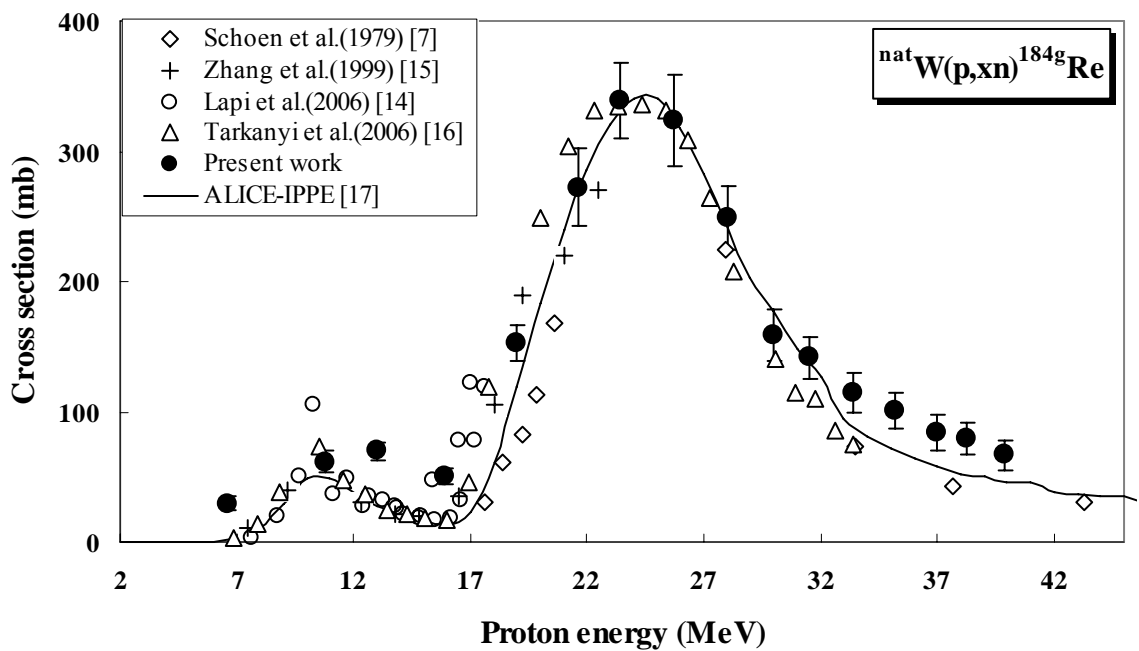

Fig. 5

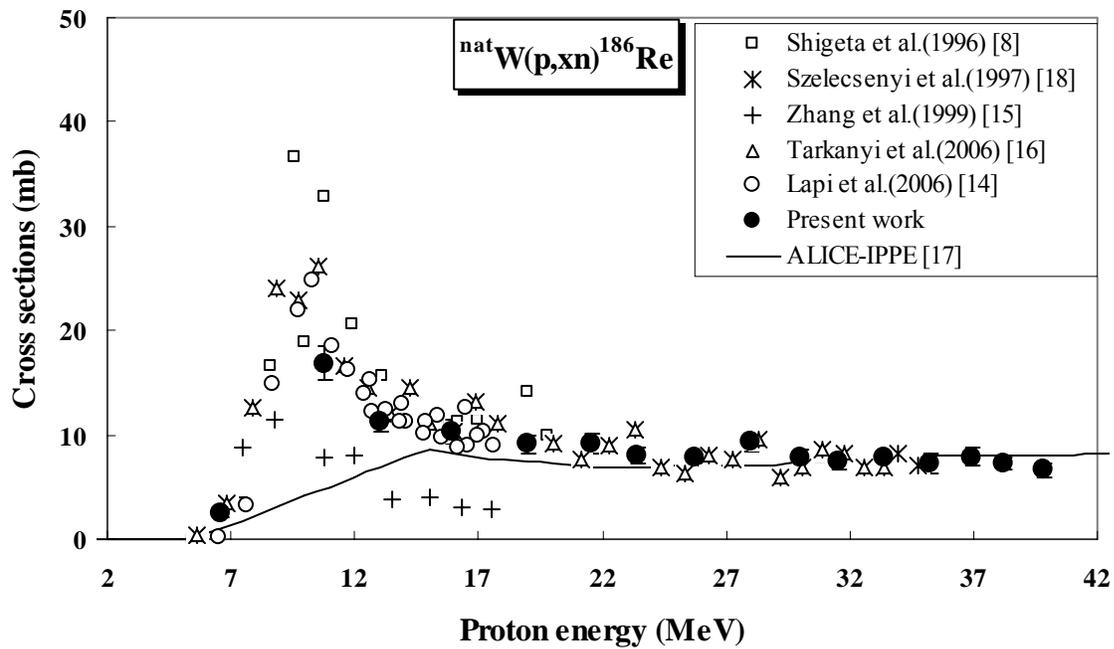

Fig. 6

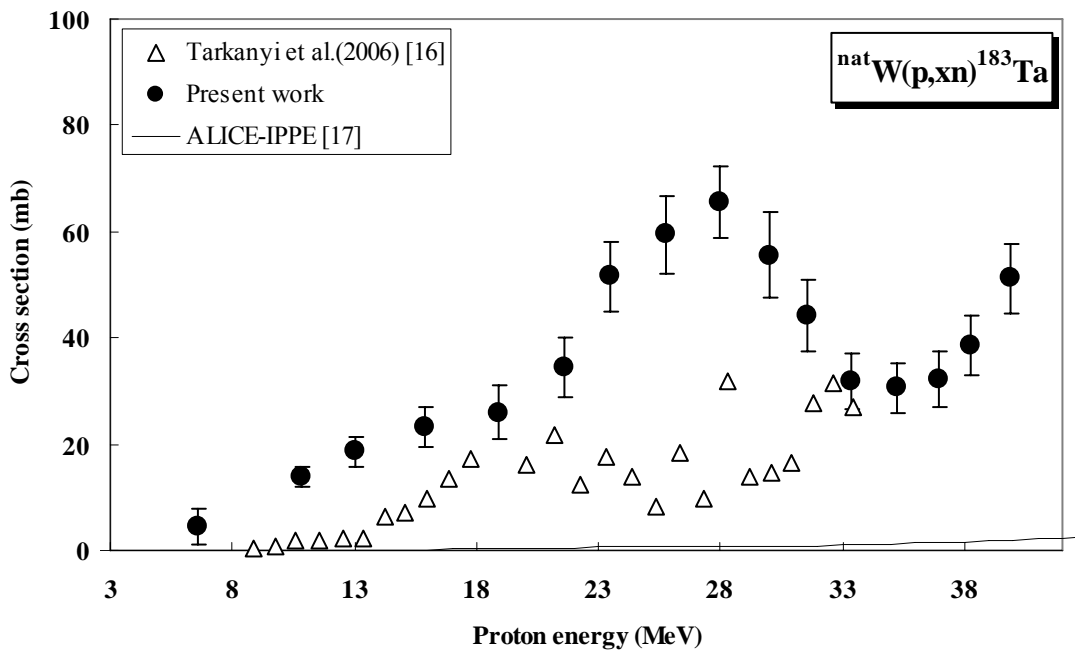

Fig. 7

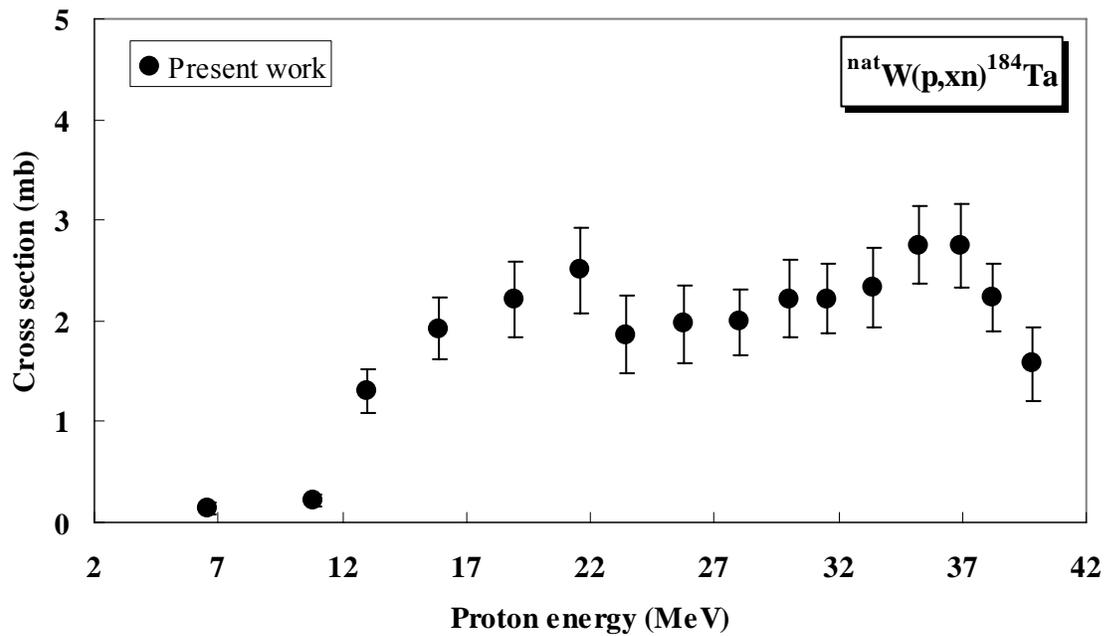

Fig. 8

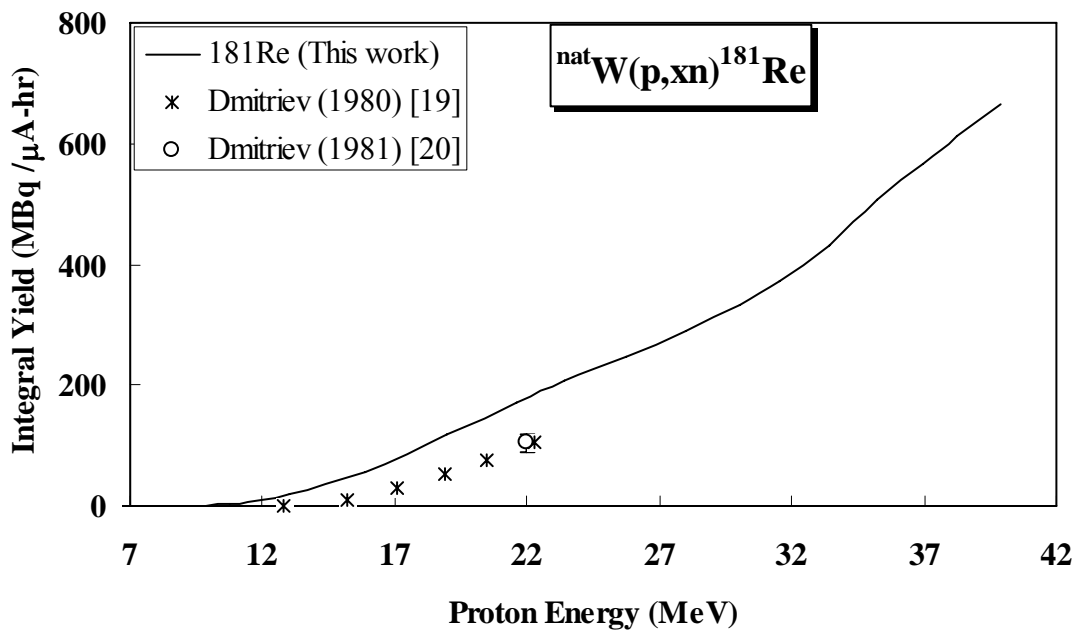

Fig. 9

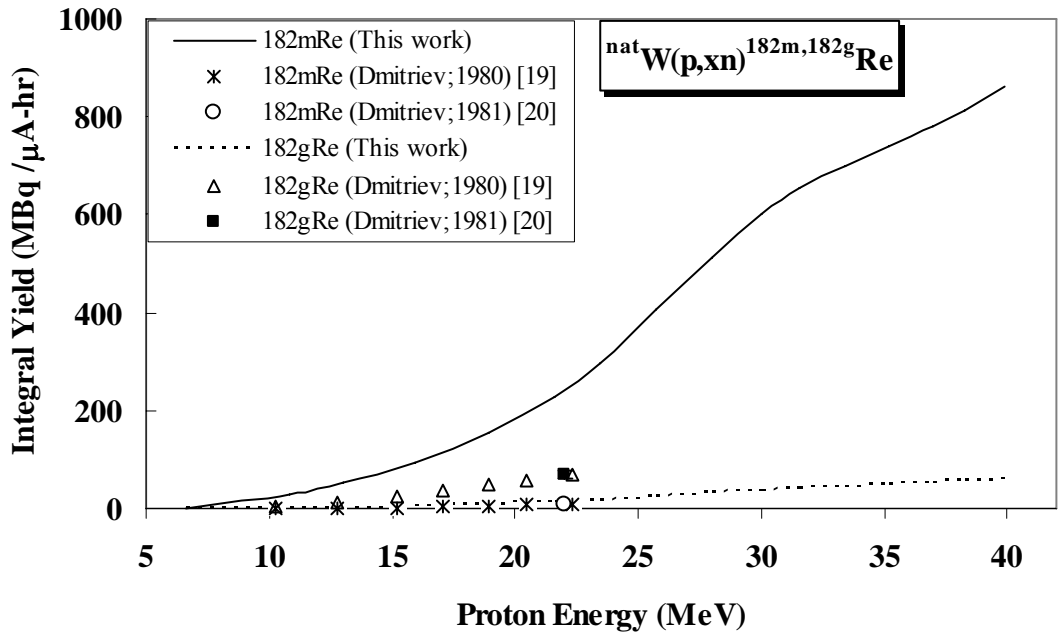

Fig. 10

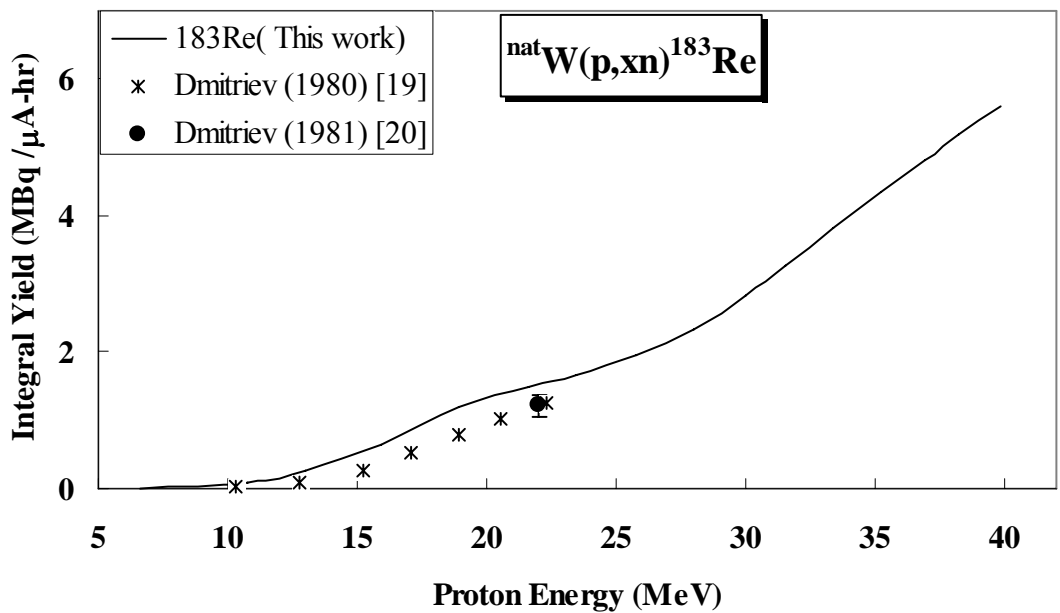

Fig. 11

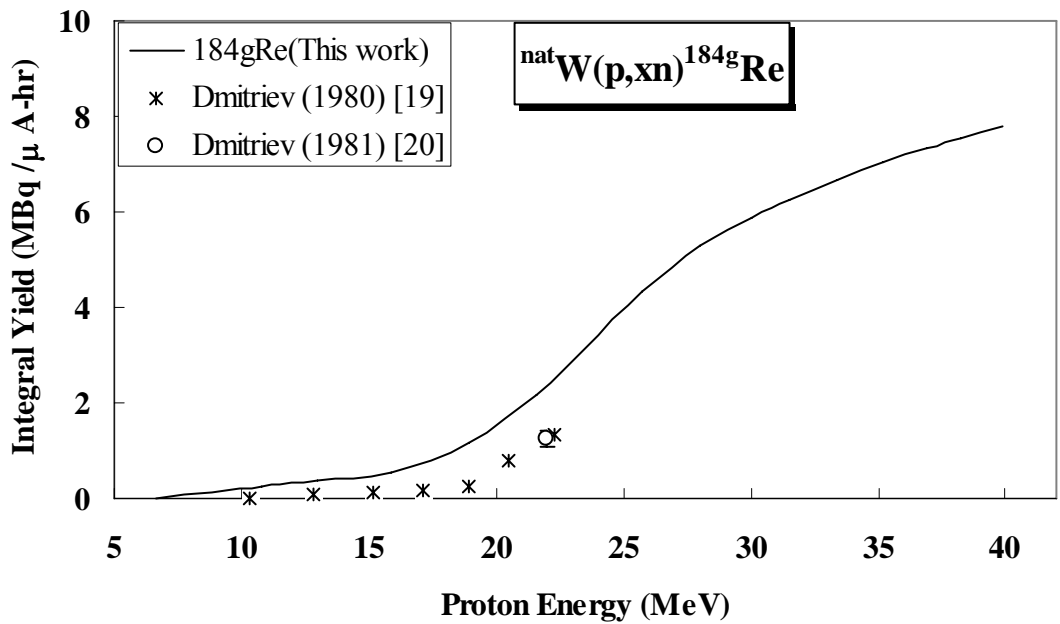

Fig. 12

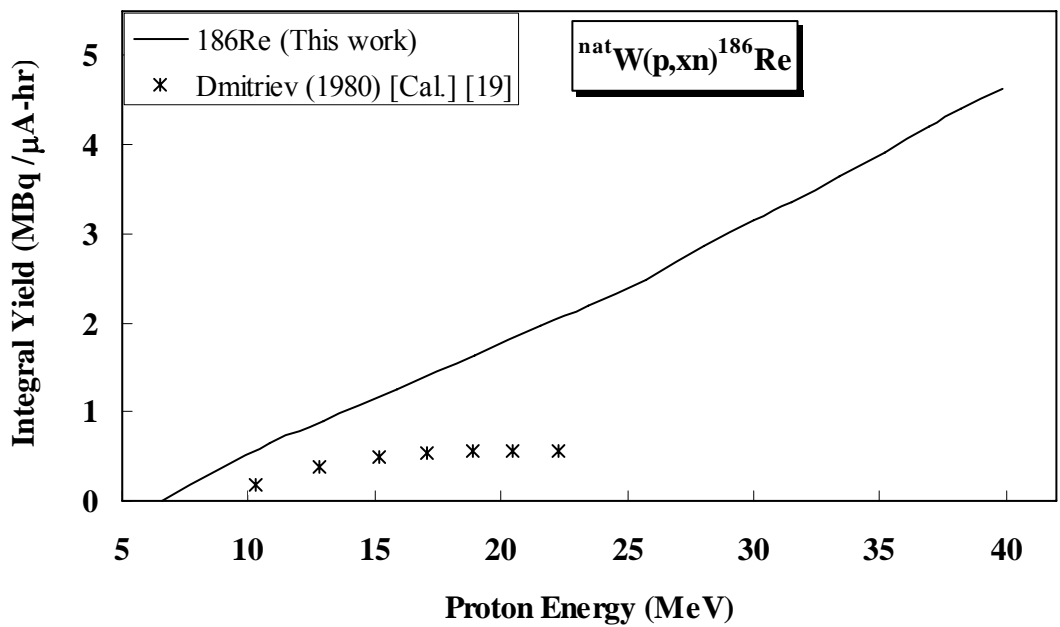

Fig. 13

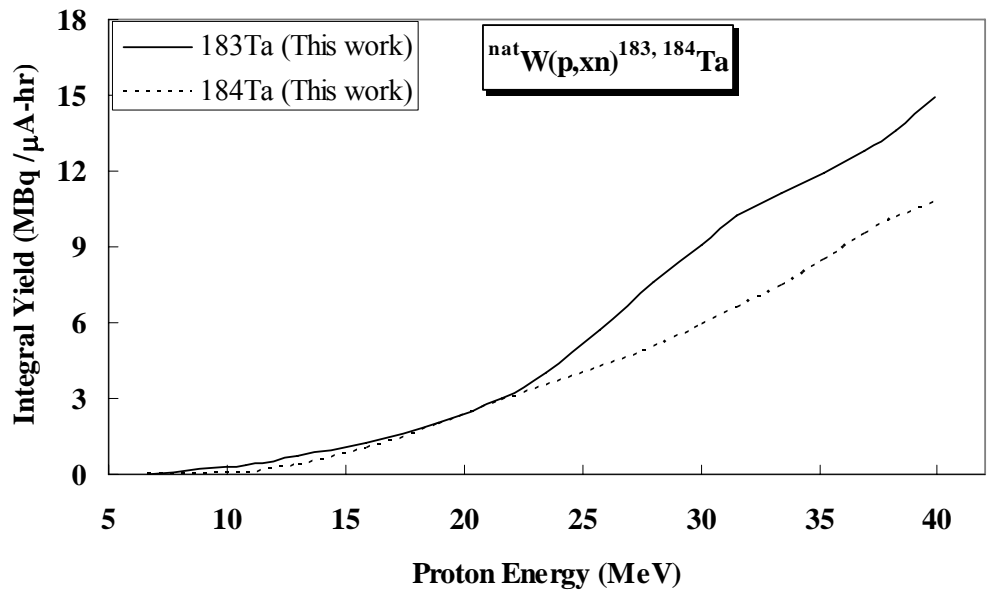

Fig. 14